\documentclass[12pt]{article}

\usepackage{graphicx}

\usepackage{amsfonts,xspace,amsmath,varioref} %

\usepackage[%
  colorlinks=false %
, pdfpagemode=None %
]{hyperref} %

\newcommand{\IC}{\mathbb{C}} %

\newcommand{\il}{\mathrm{\,iL}} %
\newcommand{\ilx}{{\mathrm{\,iL}+\mathrm{X}}} %
\renewcommand{\vref}[1]{\ref{#1}~\vpageref{#1}\unskip\xspace} %
\newcommand{\trep}{$\bar t$-rep\-re\-senta\-tion\xspace} %
\newcommand{\p}{\partial} %
\newcommand{\la}{\langle} %
\newcommand{\ra}{\rangle} %
\newcommand{\ket}[1]{\left|{#1}\right\rangle} %
\newcommand{\bra}[1]{\left\langle{#1}\right|} %
\newcommand{\sprod}[2]{\langle{#1}| {#2}\rangle} %
\newcommand{\bket}[1]{\left|{#1}\rangle\rangle\right.} %
\newcommand{\tk}{\ensuremath{t_k}\xspace} %
\renewcommand{\t}[1]{\ensuremath{t_{#1}}\xspace} %
\newcommand{\tbk}{\ensuremath{\bar t_k}\xspace} %
\newcommand{\bz}{{\bar z}} %
\newcommand{\zb}[1]{\bar z_{#1}} %

\newcommand{\CD}{\mathcal D}
\newcommand{\CE}{\mathcal E}
\newcommand{\CH}{\ensuremath{\mathcal H}\xspace}
\newcommand{\CO}{\mathcal O}
\newcommand{\CZ}{\mathcal Z}
\newcommand{\no}{\mbox{\textbf{:}}\,} 
\newcommand{\vac}{{0,0}} 

\newcommand{\betas}{{\sqrt{\beta}}}

\IfFileExists{srcltx.sty}{\usepackage[active]{srcltx}}

\def\hybrid{\topmargin 0pt      \oddsidemargin 0pt
           \headheight 0pt \headsep 0pt
       \textwidth 6.5in        
       \textheight 9in         
           \marginparwidth 0.0in
           \parskip 5pt plus 1pt   \jot = 1.5ex}

\hybrid
\numberwithin{equation}{section}

\begin{document}

\title{%
  \hfill\vbox{\normalsize\hbox{hep-th/0409129}%
    \hbox{IHES/P/04/41}%
  }\\[12pt]
Fermions in the harmonic potential and string theory}%
\author{Alexey Boyarsky\thanks{Ecole Polytechnique F\'ed\'erale de Lausanne,
    BSP-Dorigny, CH-1015, Lausanne, Switzerland}, Vadim V.
  Cheianov\thanks{NORDITA, Blegdamsvej 17, Copenhagen {\O}, DK 2100, Denmark}
  and Oleg Ruchayskiy\thanks{Institut des Hautes Etudes Scientifiques,
    Bures-sur-Yvette, F-91440, France }}
\date{} %
\maketitle
\begin{abstract}
  We explicitly derive collective field theory description for the system of
  fermions in the harmonic potential. This field theory appears to be a
  coupled system of free scalar and (modified) Liouville field.  This theory
  should be considered as an \emph{exact} bosonization of the system of
  non-relativistic fermions in the harmonic potential.  Being surprisingly
  similar to the world-sheet formulation of $c=1$ string theory, this theory
  has quite different physical features and it is conjectured to give
  space-time description of the string theory, dual to the fermions in the
  harmonic potential. A vertex operator in this theory is shown to be a field
  theoretical representation of the local fermion operator, thus describing a
  D0 brane in the string language.  Possible generalization of this result and
  its derivation for the case of $c=1$ string theory (fermions in the inverse
  harmonic potential) is discussed.
\end{abstract}

\section{Introduction and outline}
\label{sec:intro}

The $c=1$ string theory (see e.g.~\cite{moore-ginsp,klebanov-2d,Polchinski:94}
for review) is probably the simplest non-trivial representative in the vast
hierarchy of string theories.  Although it lacks many properties of its
siblings, it is still non-trivial example, admitting interesting space-time
physics.  It offers therefore very useful and already standard laboratory for
studies of general properties of string theories: open-closed string
dualities, D-branes, tachyon condensation, as well as those phenomena, which
are still hard to describe in higher-dimensional string theories (such as e.g.
black hole formation). The latter problem is of the special interest, because
two-dimensional string theory is likely to be the first model of black
hole~\cite{witten-bh,wadia-bh,wadia-colloq} which can be elaborated to such an
extent, that it would address all important questions of black hole physics,
from the initial formation to the description of final stage of Hawking
evaporation (fully taking into account quantum backreaction).

Initial progress in $c=1$ string theory was made in the early 90s (see
reviews~\cite{moore-ginsp,klebanov-2d,Polchinski:94,2dgrav-mm}) when its
``dual description'' in terms of the so-called $c=1$ \emph{matrix quantum
  mechanics} (MQM) was found and the detailed correspondence between
perturbative string theory and matrix model was developed. This allowed to use
matrix model techniques for computations of the string theory correlators in a
way alternative to the Liouville theory and provided non-trivial cross-check.
Space-time string field theory was
developed~\cite{das-jev,polchinski,wadia-bos}. The integrable structure of MQM
allows to solve the $c=1$ string theory in the large class of non-trivial
backgrounds~\cite{dijk-moore-ples,kkk,AKK}.  Recently, the interest to the
correspondence between $c=1$ string theory and MQM was revitalized.
Introduction of the D-branes as well as the progress in the boundary Liouville
CFT~\cite{zz,fzz,teshn} allowed for a new understanding of the tachyon
condensation and open-closed string duality (for a recent review see
e.g.~\cite{sen-mm}) and helped to extend the connections between MQM and
string theory to the new non-perturbative
effects~\cite{Verl1,mart,KMS,Verl2,ak-kut,c-hat,0b}.

Despite of all this progress and apparent simplicity of the model, the problem
of black hole is still unsolved~\cite{maldacena-strings2004,sen-bh}. This may
imply that some important ingredients are missing in usual set up of $c=1$
string theory. That is one reason why it is interesting to consider some
generalization of the above construction.  In this paper we will consider such
a generalization which can be both helpful in better understanding of the
relation between MQM and $c=1$ string theory and interesting by itself.

Singlet sector of MQM can be represented as non-relativistic fermions in the
inverse harmonic potential. This fact can be viewed as a reason why the model
is solvable. In terms of such fermions the most natural realization of the
integrable structure of the theory can be formulated. Moreover, recently these
fermions acquired physical interpretation~\cite{Verl1,KMS,Verl2}.  Therefore
they should be considered as fundamental objects of the theory rather than
just a technical tool. So it is useful to consider generalizations of the
theory which are the most natural from the point of view of the fermions and
not only from the world-sheet string theory perspective.

The simplest modification of this kind is to consider the fermions in the
harmonic (rather than inverse harmonic) potential.
\begin{figure}[t]
  \centering \includegraphics[width=0.3\textwidth]{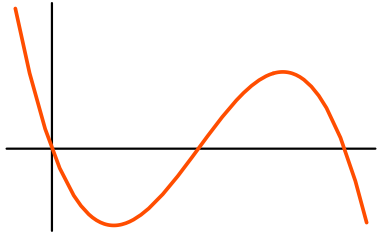}
  \caption{MQM in such potential has two universal regions -- harmonic
    potential near its minimum and inverse harmonic near its maximum.}
  \label{fig:potential}
\end{figure}
This system has several interesting features. First of all, it has a discrete
spectrum which, for example, makes much clearer the constructions of
many-particle wave functions. This is important e.g. when one is interested in
non-perturbative effects. Another important feature is the finite depth Fermi
sea instead of infinite Dirac sea, which is natural for
\emph{non-relativistic} fermions. These features do not necessary imply that
the theory has no dual string theory description. One can imagine the theory
in, say, potential sketched on the Figure~\ref{fig:potential}. Such a theory
looks like inverse harmonic oscillator near the maximum (apart from the fact
that all states are metastable) and almost like harmonic oscillator near the
minimum. The complete theory in such a potential can be considered from the
point of view of the inverse harmonic potential as one of the possible
(non-universal) regularizations.  Both limits are universal and are related by
the analytic continuation (see~\cite{akk2,c1} for the discussion).  In the
first limit the (double)scaled theory has infinite Fermi sea and continuous
spectrum.  This corresponds to the continuous limit of MQM from the point of
view of surface triangulation and gives $c=1$ string theory in the linear
dilaton background.  If one is instead interested in some other string theory,
this feature is not necessary.  Moreover, it can in principle even prevent one
from describing in terms of the fermions a theory which in fact does not imply
infinite Fermi sea for fermions (D-branes) whose condensate is a dual
description of its background.

\subsection*{Outline of the results}
\label{sec:outline}

In this paper we take the following approach. We try to find a bosonic theory
which is equivalent to the system of fermions in the harmonic potential in a
sense that all operators and their expectation values are in one to one
correspondence.  This correspondence is directly \emph{derived} and can be
viewed as an example of an \emph{exact bosonization} of non-relativistic
fermions.\footnote{The exact bosonization for the case of inverse harmonic
  potential was first discussed
  in~\cite{wadia-bos,wadia-w-inf,wadia-time-dep}, where it was realized in
  terms of quantum phase space distribution of the Fermi liquid
  (see~\cite{wadia-tachyon} for a recent results and applications to the study
  of tachyon dynamics in $c=1$ string theory). This approach is somewhat
  similar to the one, undertaken in this paper. We leave detailed comparison
  for the future work.}

This derivation has two steps. First, one tries to bosonize fermions in terms
of free scalar using the analogy with the theory of relativistic fermions.  In
the case of the inverse harmonic potential this gives an interacting and
highly non-trivial effective field theory~\cite{das-jev,polchinski,wadia-bos}
which simplifies only asymptotically, far from a classical turning point,
where fermions are almost relativistic.

Many-particle state of the fermions form a Fermi sea in the phase space of the
system.  The oscillators of this bosonic effective theory correspond to the
small deformations propagating along the Fermi sea. These deformations are
well-defined in the phase space and are two-dimensional in their nature.  It
means, in particular, that the wave function, corresponding to the perturbed
Fermi sea, are not eigen-functions of the Hamiltonian with the harmonic
(inverse harmonic) potential. If one tries to find a Hamiltonian for which
this wave-function will be an eigen-state, such Hamiltonian turns out to be
very complicated and non-local (c.f.~\cite{AKK,c1}).\footnote{The system is
  integrable and this non-local operator is related to the original
  Hamiltonian by the \emph{dressing} procedure in the terminology of
  integrable systems.}

However, in both cases of harmonic and inverse harmonic potential there exist
a very simple equivalent reformulation of the problem in terms of a system of
free fermions in two dimensions in the strong uniform magnetic
field~\cite{c1}. In such field the system is confined to the zero energy level
of its Hamiltonian. This level is infinitely degenerate.  It was shown
in~\cite{c1} that all wave-functions which correspond to all possible
incompressible deformations of the Fermi sea belong to this level. The
equivalence with the one-dimensional problem can be established as follows. To
consider thermodynamic properties of the system, one needs to choose an
$N$-particle ground state.  According to the Fermi statistics, it is usually
build as a Slater determinant of the first $N$ eigen functions of the
Hamiltonian. When working with the space of the solutions of the equation
$H_{2d}=0$, to do this one needs to choose some ordering on this infinitely
degenerate energy level.  This is done by choosing any (Hermitian) operator
commuting with the Hamiltonian and choosing ordered basis of its
eigen-functions. If one chooses the angular momentum operator $L$ for the
ordering, the system of equations
\begin{equation}
  \label{eq:220}
 H_{2d} \psi_E(z_+,z_-)=0, \qquad L \psi_E(z_+,z_-)= E \psi_E(z_+,z_-) 
\end{equation}
is exactly equivalent to the one-dimensional equation $H_{1d} \psi_E(z)= E
\psi_E(z)$ and the $N$-particle ground state of this system coincides with
undeformed Fermi sea in the phase space of one-dimensional system.  If one
chooses some other operator for the ordering, the ground state and, therefore,
the shape of the Fermi sea will be different. But the wave-functions
corresponding to all of them have one common property -- they satisfy first of
the equations~(\ref{eq:220}). That is why two dimensional representation is
convenient for the study of effective field theory description of the system.

As a result, instead of eigen-value problem of one-dimensional Hamiltonian of
MQM one has a Wheeler-DeWitt type equation saying that the two-dimensional
Hamiltonian is equal to zero. The same happens if one relates e.g.
quantization of the relativistic particle in terms of coordinate time (after
fixing some space-time splitting) to the covariant quantization of the same
system with respect to its proper time done without any coordinate choice.
This is why such a two dimensional reformulation can be considered to be a
relativistic reformulation of the MQM system~\cite{c1}.  In this
two-dimensional formulation different shapes of the Fermi sea on the MQM side
all become solutions of the same two-dimensional Hamiltonian (Wheeler-DeWitt)
equation, corresponding to the different choices of the space-time splitting.

This reformulation allows to write down a bosonization of the system in terms
of the free chiral two-dimensional boson. In case of inverse harmonic
potential, this result is contained, somewhat implicitly, in~\cite{AKK} and
for both potentials in~\cite{c1}.

The bosonization described above has nevertheless one unpleasant feature.  The
theory of free boson obviously has ultra-violet divergences.  On the other
hand, the initial theory of $N$ non-relativistic fermions is finite at any
scale. The reason for this discrepancy is that the system of such fermions has
the finite number of degrees of freedom.  Therefore, not all degrees of
freedom of bosonic field are independent for finite $N$. To be able to
reconstruct properly quantum-mechanical correlators for a finite, but large
$N$ from the field theoretical description, one needs to be more careful with
the large $N$ limit.

The solution of this problem, described in the present paper for the case of
the harmonic potential, gives rise to the following picture. Instead of chiral
boson, we have the system of the full boson $X$ plus (modified) Liouville
field with the action
\begin{equation}
  \label{eq:15}
   S_\ilx[X,\phi] =\frac{1}{2 \pi} \int d^2 z \left\{\partial X \bar \partial X
    +  \partial \phi \bar \partial \phi+ \frac{2i}h\phi -
    \mu e^{i \phi}
  \right\}
\end{equation}
(where $h$ is a constant of order $1/N$).  Vertex operator in this system,
corresponding to the local fermion operator on the quantum mechanical side,
looks like
\begin{displaymath}
\label{vertex}
  \psi^+_\il(\zeta)=e^{\frac {i}2
    \left(\phi(\zeta,\bar\zeta)+X(\zeta)-\bar X(\bar\zeta)\right)}
\end{displaymath}
Let us stress, that this theory is
an \emph{exact} bosonization of the system of fermions in the harmonic
potential and it is derived explicitly in the present paper.  In the limit
when the Fermi sea is a step-function, Liouville field freezes to its
background value and effectively imposes the Dirichlet boundary condition on
the boson $X$. In this way we recover the initial bosonization in terms of
free chiral boson.

Surprisingly, this system looks very much like string theory. There are,
however, two differences with the string theory in the linear dilaton
background~\cite{Polchinski:94,klebanov-2d,moore-ginsp}. First, the exponent
in the Liouville field is imaginary. This is due to compactness of the Fermi
sea. This property can be removed by analytic continuation, which should be
done simultaneously in the $z$-plane and in the space of the field $\phi$.
This continuation is tricky. For the chiral boson without Liouville field
analytic continuation in the $z$-plane relates the theories in harmonic and
inverse harmonic potentials as well as bosonization formulas for these
theories~\cite{c1}. Generalization of this analytic continuation for the full
theories is discussed in the section~\ref{sec:2dstr}.

Another important difference is the appearance of the background term
in~(\ref{eq:15}).  This background is not a usual curvature term as in the
action of $c=1$ string theory and thus cannot be removed from the action by a
shift of field $\phi$. Indeed, in the $c=1$ string theory the field
redefinition $\phi\to\phi+ \sigma$ can be compensated by the Weyl rescaling of
the world-sheet metric $g_{ab} \to g_{ab}e^{2\sigma}$. Thus, the curvature
term can be eliminated and the Liouville term $\sqrt g\mu e^\phi$ remains
unchanged. In the case involved, we do not have an analog of the $\sqrt g$
term in the action~(\ref{eq:15}). As a result, the field redefinition $i\phi
\to i\phi - \frac{|z|^2}h$ would modify the Liouville potential
$\mu\,e^{i\phi} \to \mu\,e^{i\phi - \frac{|z|^2}h}$ (we will return to this
question in Sections~\ref{sec:liouville-eq}, \ref{sec:2dstr}). Thus, this
linear term is a real modification of the theory and it would be there also in
the same analysis for the inverse harmonic potential. The result of this
modification is important -- it is responsible for the regular behavior of the
correlators at small scales, as discussed in the
Sections~\ref{sec:liouville},~\ref{sec:density-matrix}.  This feature, which
is the main reason why Liouville field is necessary in the bosonization of the
non-relativistic fermions, is expected from the \emph{space-time} picture of a
string theory. Indeed, our analysis is the analog of the Das-Jevicki
collective field theory approach (made for harmonic potential, in a different
way and valid at any scale).  This collective theory is believed to be indeed
the space-time theory.  Therefore our theory should have the same
interpretation as well.  This space-time interpretation and why our theory is
so similar to the world-sheet theory is not yet clear for us.\footnote{The
  fact that the fermions in the harmonic potential have dual string theory
  description was also noticed in~\cite{berenstein}. When this paper was at
  the final stage of preparation, the work~\cite{izh-mcg} appeared, where the
  theory of ``imaginary'' Liouville field coupled to the free boson was
  claimed to be dual to the matrix model in the harmonic potential.  In our
  paper we give a direct (and independent) prove of a similar fact -- namely,
  that the system of free fermions in the harmonic potential is equivalent to
  the system~(\ref{eq:15}) which is almost (but not completely!) the same as
  the one of~\cite{izh-mcg}.  One important difference between our theory and
  that of~\cite{izh-mcg} is that in the latter background term is treated as a
  usual curved metric background in Liouville theory. Thus it can be removed
  by a field redefinition.  In our case it is not true, as it was discussed
  above as well as in the end of Section~\ref{sec:liouville-eq}. The
  importance of this modification is that it makes correlators finite at small
  scales.  This makes two theories physically different. Due to the same
  reason, ``duality map'' seems to be different as well. For example, local
  fermion operators correspond to the vertex operators on the ``string
  theory'' side in our case.  The two-point function of vertex operators is
  non-singular, in agreement with the quantum mechanical result. At the same
  time if one calculates two point function of the same vertex operators in
  the model suggested by~\cite{izh-mcg}, it will have an ultra-violet
  singularity usual for a conformal field theory. It is therefore unclear how
  to build the operators representing fermions correctly in the
  model~\cite{izh-mcg}.  Unfortunately, in~\cite{izh-mcg} the space-time
  interpretation of the theory is not discussed. If the proposed there duality
  is correct, it gives a world-sheet rather than a space-time theory.}

\section{Collective field theory for the fermion density}
\label{sec:liouville}

We study one-dimensional fermions in the harmonic potential. As it was
discussed in~\cite{c1} and reviewed in Section~\ref{sec:intro}, this theory
can be equivalently reformulated as the system of two-dimensional fermions in
the magnetic field.  In this section we will show that it is equivalent to the
``imaginary'' Liouville field theory. This field theory is in its spirit a
collective field theory~\cite{jev-sakita,das-jev}, written in terms of the
density of fermions, rather than describing the dynamics of individual
fermions.

We begin with the following family of ground states of the system of
fermions:\footnote{We introduces an additional parameter $\beta=1, 3, 5, \dots
  $ In case $\beta=1$ (free fermions in the magnetic field) the wave
  function~(\ref{Laughlin}) is a Slater determinant of the first $N$ monomials
  $\Psi_0 \sim \det (z^{i-1}_j)$, which is just the ground state of the system
  of $N$ non-interacting fermions, ordered with respect to their angular
  momentum. Case $\beta>1$ corresponds to a more complicated interacting
  system: it is a celebrated Laughlin's wave-function~\cite{laughlin},
  describing fractional quantum Hall state with the filling fraction
  $\beta^{-1}$. Naively, case $\beta>1$ does not correspond to the singlet
  sector of MQM and thus has no immediate connection to our problem.  However,
  surprisingly, all the derivations work exactly the same for any $\beta$, so
  we can keep it ``at no cost''. Moreover, as we will see later
  (Section~\ref{sec:idls}), $\beta$ plays the role of the radius of
  compactification for one of the fields in the collective theory of the
  system~(\ref{Laughlin}).
  
  It is interesting to note that the theory with $\beta>1$ can arise in the
  non-singlet sectors of the MQM~\cite{boul-kaz}.} 
\begin{equation}
  \Psi_0(z_1,\dots, z_N)\equiv\langle z_1,\dots , z_N|
  0\rangle=\frac{1}{\sqrt N!} \prod_{i<j}^{N}(z_j-z_i)^{\beta}
  e^{-\frac{\beta}{2} \sum\limits_{j=1}^N
    |z_j|^2}
\label{Laughlin}
\end{equation}
\begin{figure}[t]
  \begin{center}
    \includegraphics[width=0.35\textwidth]{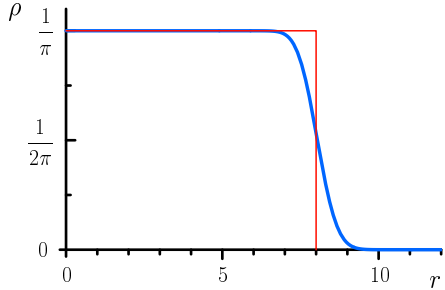}
\caption{Typical density distribution as a smoothed out
  step-function. The plot is shown for $N=64$ particles.}
\label{fig:density} 
\end{center}
\end{figure}
The discussion below will be concerned with the statistically large number of
particles.  The thermodynamic limit of this system is defined as $N\to
\infty$. It is well known~\cite{laughlin,qhe-book} that in this limit ground
state~(\ref{Laughlin}) describes fermions forming round droplet of radius
$\sqrt{N}$ with the constant density equal $\frac1\pi$ inside and zero
outside.  Namely, the average of density operator $\rho(|z|)=\bra0\rho(z,\bar
z)\ket0$ has the profile, shown on Fig.~\vref{fig:density}.  Notice that the
width of the ``edge'' (transition region between $\rho=\mathrm{const}$ and
$\rho=0$) is proportional to 1 in our units (and is small compared to the
radius). In many cases it is enough to approximate this profile by the
step-function: $\rho(|z|) =\frac1\pi\theta(\sqrt N - |z|)$, however, one
should take into account, that this is a \emph{non-uniform approximation}: for
$|z-\sqrt N|\ll 1$ difference between the real distribution and the
step-function is always of order one.

Wave-function~(\ref{Laughlin}) is \emph{not} normalized. The norm of the
ground state~(\ref{Laughlin}) is defined by
\begin{equation}
  \tau_N=\int d^2 z_1 \dots d^2 z_N | \Psi_0
  |^2
  \label{ZN}
\end{equation}
$\tau_N$ can be interpreted as a partition sum of the Coulomb plasma, by
noticing that eq.~(\ref{ZN}) can be written as
\begin{equation}
 Z_N(\beta)=\int d^2 z_1 \dots d^2 z_N | \Psi_0 |^2 = \int_\IC d^2 z_1
 \dots d^2 z_N\, 
 e^{-\beta \CE_c(z_1, \dots , z_N) } 
\label{PF}
\end{equation}
where
\begin{equation}
\CE_c(z_1, \dots, z_N)=-\sum_{i<j}{2}\ln| z_i-z_j |
+\sum_{i=1}^N  | z_i|^2
\label{coulombH}
\end{equation}
is the classical energy of a 2D Coulomb plasma of particles of charge $e^2=2$
moving in the uniform background of particle density $\rho_0=\frac 1\pi$. This
Coulomb plasma analogy allows for a simple and transparent way to understand
what kind of collective field theory one should expect.  First of all, it is
easy to show that the density distribution has be approximated by the
step-function (see e.g.~\cite{laughlin,qhe-book,wieg-qhe,prb} in the context
of quantum Hall effect or~\cite{kkmwz,wz-nmm} in the context of matrix
models). For many problems step-function-like density is enough, but let us
stress once again that it is a non-uniform approximation on the edge of the
droplet.

This approximation can be significantly improved by noticing that~(\ref{PF})
looks like a plasma with the temperature $\beta^{-1}$.  Indeed, let us
introduce the electric potential $\varphi(z, \bar z),$ satisfying the 2D
Poisson equation
\begin{equation}
  \Delta \varphi = 4 \pi e \rho(z, \bar z)
  \label{Poisson}
\end{equation}
where $\rho = \rho_p - \rho_0$ is the total charge density created by both
particles and background in the system. The key approximation, often used for
such a system is to say, that the distribution of the mobile charge in the
external electric field is given by the Boltzmann law (recall that the
temperature is $\beta^{-1}$)
\begin{equation}
  \rho_p(z, \bar z)= \mu e^{ -\beta \,e  \varphi(z,\bz)}
  \label{Boltzmann}
\end{equation}
where $\mu$ is the \emph{fugacity}.  Combining equations (\ref{Boltzmann}) and
(\ref{Poisson}) we find that the scalar potential $\varphi$ satisfies the
equation
\begin{equation}
\Delta\varphi = 4 \pi e \left(\mu e^{-\beta e \varphi(z,\bz)}- \rho_0\right)
\label{Liouville}
\end{equation}
For our purposes the parameter $\mu$ is to be chosen so that the total number
of particles be equal to $N$
\begin{equation}
  \mu \int d^2 z\; e^{ -\beta\,e \varphi(z,\bz)} = N
  \label{constraint}
\end{equation}
Equations~(\ref{Liouville}) and~(\ref{constraint}) constitute the closed set
of equations which determine the density distribution of the fermionic droplet
in our approximation.\footnote{This equation can be derived from the Ward
  identities in the context of normal and complex matrix models~\cite{wz-nmm},
  which is, of course, closely related to this problem.} 
Solution to this equation gives rise to a smooth function, which serves as a
better approximation to the real density, computed over~(\ref{Laughlin}). Loop
expansion of the field theory (discussed in the Section~\ref{sec:tk}) will
give a uniform approximation.  This smoothing out is responsible for getting
rid of the ultra-violet singularities in the fermion correlators (c.f.
Section~\ref{sec:density-matrix}).

Eq.~(\ref{Liouville}) is clearly of the Liouville type (modulo the constant
term).  Thus, we can expect, that written in terms of field $\varphi$ (which
is, essentially, the logarithm of the fermion density), one should get the
collective field theory, related to that of Liouville.

\subsection{The Liouville field}
\label{sec:liouville-eq}

To derive a field theoretical reformulation of the quantum mechanics of $N$
fermions~(\ref{Laughlin}), we will show that in the large $N$ limit the grand
partition function of the system~(\ref{coulombH}) (or,
equivalently,~(\ref{Laughlin})) can be calculated by means of
field-theoretical path integral.\footnote{Such approach was first used
  in~\cite{polyakov} to obtain a diagram technique for a description of phase
  transitions in various systems.} 
The saddle point equation of this theory will reproduce~(\ref{Liouville}).

We define grand canonical ensemble of the Coulomb plasma via\footnote{As
  $\tau_N \sim e^{N^2\log N}$ for $N\to\infty$, formal sum~(\ref{eq:22}) is
  divergent.  This, however, can be cured by an IR regularization. For
  details, see Appendix~\ref{sec:careful} and work~\cite{liouv-big}.
  \label{fn:1}}
\begin{equation}
  \label{eq:22}
  \CZ(\xi)  \equiv \sum_{N=0}^\infty \frac{\xi^N}{N!}\tau_N=\sum_{N=0}^\infty
  \frac{\xi^N}{N!}Z_N(\beta)=\sum_{N=0}^\infty 
  \frac{\xi^N}{N!} \int d^2 z_1 \dots d^2 z_N |\Psi_0(z_1,\dots,z_N)|^2
\end{equation}
The first step towards the field-theoretical reformulation is to rewrite the
Hamiltonian~(\ref{coulombH}) in terms of a correlator in a free boson field
theory:
\begin{equation}
  e^{-\beta\CE_c(z_1, \dots, z_N)} =\left \langle \prod_{j=1}^N e^{i \phi(z_j,
      \bar z_j)-\beta | z_j|^2} \right \rangle_{S_0}
  \label{HFT}
\end{equation}
where the average
\begin{equation}
  \langle  \dots \rangle_{S_0} =
  \frac{1}{Z_{\phi}} \int \mathcal D \phi   \dots  e^{-S_0(\phi)},\qquad
  Z_\phi = \la 1\ra_{S_0}
\end{equation}
is taken with respect to the Euclidean Gaussian action of a free massless
boson\footnote{One has to be careful with the presence of both UV and IR
  divergences in this action. We will ignore all these subtleties for now,
  referring reader to the appendix and to the future work~\cite{liouv-big}.}
\begin{equation}
  S_0[\phi]=
  \frac{1}{2 \pi \beta} \int d^2 z
  \,\partial \phi \bar \partial \phi
  \label{S0}
\end{equation}
with both short and long-distance cutoffs implied (see
Appendix~\ref{sec:careful}).  For the normalization $\tau_N$ (eq.~(\ref{ZN}))
this representation gives
\begin{equation}
  \tau_N=\int d^2
  z_1\dots d^2 z_N \left \langle \prod_{j=1}^N e^{i \phi(z_j, \bar
      z_j)-\beta | z_j|^2} \right \rangle_{S_0}
  \label{ZNFT}
\end{equation}
Consider now the grand partition function~(\ref{eq:22}). Substituting the
representation~(\ref{ZNFT}) in it, we find
\begin{equation}
  \CZ(\xi)=\frac{1}{Z_{\phi}} \sum_{N=0}^{\infty} \frac{\xi^N}{N!}  \int \mathcal D
  \phi \,e^{-S_0} \left[\int d^2 z\, e^{i \phi
      -\beta | z|^2 }\right]^N \label{Zsum}
\end{equation}
The sum over $N$ is easily performed yielding
\begin{equation}
\CZ(\xi) = \frac{1}{Z_\phi} \int \mathcal D \phi\, e^{-S_{iL}[\phi]},
\label{Zint}
\end{equation}
where
\begin{equation}
  S_{\il}[\phi]= \frac{1}{2 \pi \beta} \int d^2 z \left[\partial \phi \bar
    \partial \phi -\mu e^{i \phi- {\beta}|z|^2 }\right] \label{SLi}
\end{equation}
Equations~(\ref{Zint}) and~(\ref{SLi}) define the path integral representation
for $\CZ$. The coupling constant
\begin{equation}
  \mu = {2\pi  \xi\beta}
 \label{eq:26}
\end{equation}
has the dimension of inverse length squared.\footnote{$\mu$ is a running
  coupling constant satisfying the RG equation $\frac{d \log\mu}{d\ln
    \Lambda}= \beta$. See Appendix~\ref{sec:careful} for details.}

We have shown that the grand canonical ensemble of $N$ fermions can be
rewritten as the field theory~(\ref{SLi}).  Several comments are of order
here. First, we have derived the theory, similar to the usual Liouville field
theory, however with an imaginary (rather than real) exponential term --
\emph{Liouville theory with imaginary coupling}. For the sake of shortness we
will often call it \emph{imaginary Liouville theory} and sometimes simply
Liouville theory, whenever it does not cause a confusion.

As already discussed in Introduction, this action has an apparent distinction
from the ordinary Liouville theory. Namely there is a background term
$e^{-\beta|z|^2}$.  Notice, that by the field redefinition $\phi\to \phi -
i\beta|z|^2$ this term can be traded for the linear term in the action of the
form $2i\beta\phi$ (as in~(\ref{eq:15})), which resembles a background
curvature term in the usual Liouville action:
\begin{equation}
  \label{eq:37}
  S_L[\phi] = \frac 1{8\pi}\int d^2 z\,\sqrt g\left(\p \phi \bar\p \phi + R(g)
    \phi +\mu e^\phi\right)
\end{equation}
Yet, it is \emph{not} a curvature (unlike that in~(\ref{eq:37})). Indeed, in
the action~(\ref{eq:37}) one can perform the shift of variable $\phi\to
\phi+\sigma$, such that $R-2\nabla^2\sigma = 0$. Then the curvature term would
disappear, while Liouville terms would remain the same, as an additional
factor $e^\sigma$ would be absorbed into the Weyl rescaling of the metric
$g_{ab} \to g_{ab} e^{2\sigma}$. On the other hand, in the imaginary Liouville
action~(\ref{SLi}) there is no analog of the Weyl symmetry and thus the
background term constitutes the physical modification of the
action~(\ref{SLi}). This term is responsible for making short-distance
behavior of the correlators, computed in the theory~(\ref{SLi}) finite.  We
will discuss the importance of this term, as well as other (dis)similarities
with the usual $c=1$ string theory in Section~\ref{sec:2dstr}.

Notice also, that the problem, mentioned in the footnote~\vref{fn:1},
translates in the language of the theory~(\ref{SLi}) into the well-known fact
that one does not have an analytic expansion in parameter $\mu$ (as could be
erroneously concluded from the definition~(\ref{eq:22})).

At last, it can be shown (if one takes care of all the subtleties, which arose
in the derivation of Section~\ref{sec:liouville-eq}, c.f.
Appendix~\ref{sec:careful}) the correlators computed with respect to the grand
canonical and canonical ensembles with the number of particles given by $\bar
N=\la N\ra_{GCE}$ will be the same.

\subsection{Operators in the imaginary Liouville theory}
\label{sec:dens-op-liouv}
 
Field theory~(\ref{SLi}) should be an equivalent description of the original
system of fermions in the magnetic field with the ground
state~(\ref{Laughlin}). Thus, we need to identify the vertex operators in this
theory, whose correlators reproduce the quantum-mechanical averages of the
fermions.

We begin with the fermion density operator, given in the first-quantized
formalism by
\begin{equation}
  \label{eq:28}
  \bra\Psi\rho(\zeta)\ket\Psi= \int d^2 z_{1} \dots 
  d^2 z_N \,|\Psi(z_1, \dots,  z_N,\,\zeta)|^2
\end{equation}
On the ground state this operator acts as
\begin{equation}
  \label{eq:46}
  \bra0\rho(\zeta)\ket0= e^{-\beta|\zeta|^2}\int d^2 z_{1} \dots 
  d^2 z_N \,\prod_{j=1}^N|\zeta-z_j|^{2\beta}|\Psi_0(z_1, \dots,  z_N)|^2
\end{equation}
To find a path integral representation of the density operator~(\ref{eq:28})
we need to pass to the grand canonical ensemble in the correlator $ \bra 0 \no
\rho(z_1)\dots \rho(z_n) \no \ket0$:
\begin{equation}
  \bra 0  \no \rho(z_1)\dots \rho(z_n) \no \ket 0
  =\frac{1}{\CZ(\mu)} \sum_{N=n}^{\infty} \frac{\xi^N}{N!}\frac{N!}{(N-n)!}
  \frac{1}{\tau_N} \int
  d^2 z_{n+1} \dots d^2 z_N\, e^{-\beta \CE_c(z_1, \dots, z_N)}
\label{eq:25}
\end{equation}
Substituting the exponential from~(\ref{eq:25}) with eq.~(\ref{HFT}) and
performing the sum we obtain the $n$-point density matrix in terms of the
average in imaginary Liouville theory:
\begin{equation}
  \bra0 \no \rho(z_1)\dots \rho(z_n) \no\ket0  =  \xi^{n}\left
    \langle \prod_{j=1}^{n} e^{i\phi(z_j, \bar
      z_j)-\beta|z_j|^2}\right\rangle_{S_{\il}}
  \label{denmatLio} 
\end{equation}
where the average is taken in the theory with the action (\ref{SLi}).
Comparing the right hand side and the left hand side of eq.~(\ref{denmatLio})
we identify the density operator in the Liouville theory:
\begin{equation}
\rho(z, \bar z)=\frac{\mu}{2\pi\beta} e^{i\phi(z,\bar z) -\beta| z|^2}
\label{rhoLio}
\end{equation}
where $\mu$ is defined in eq.~(\ref{eq:26}).

Thus we can reproduce in the field theoretical formulation density-density
correlation functions of the quantum mechanics of fermions. In particular,
saddle point equation of~(\ref{SLi}) (which coincides with~(\ref{Liouville})
as expected) reproduces qualitatively the correct behavior of the density.
Indeed, it can be shown (see~\cite{liouv-big}) that it looks as a step
function smoothed out at the edge (see qualitative picture on the
fig.~\vref{fig:density}).  Other properties of the density correlators will be
discussed later.  Here we would like to stress that to have the full
correspondence between the quantum mechanics and Liouville theory we are still
lacking at least one important ingredient.  Namely, an important object in
quantum mechanics of the system of many fermions is a \emph{local fermion
  operator}.  For example, instead of the density $\rho(z,\bz)$ we could
calculate fermionic two point function $\la\psi^+ (z) \psi
(z')\ra$.\footnote{Of course $\la\psi^+ (z) \psi (z')\ra\to\rho(z,\bar
  z)$ as $z'\to \bz$.} 
Important feature of the fermion operator is that it is \emph{chiral}.  In the
first quantized formalism the operator $\psi^+(\zeta)$ multiplies the wave
function by $(z_i-\zeta)$:\footnote{We will omit $\beta$ in the rest of this
  section to make presentation more transparent. Full derivation will be
  presented in the section~\ref{sec:free-boson}.}
\begin{equation}
  \psi^+_{QM} (\zeta) \Psi_0(z_1\dots z_N)= \Psi_0(z_1\dots z_N,
  \zeta)=\prod_{i=1}^N (z_i-\zeta) 
  \Psi_0(z_1\dots z_N)
  \label{eq:47}
\end{equation}
The only ingredient at our disposal so far -- the Liouville field
$\phi(z,\bz)$ -- is not chiral: insertion of density operator $\exp(i\phi)$
corresponds to the multiplication of the ground state by the non-chiral object
$\prod\limits_i
|z_i-\zeta|^{2}$ (c.f.~(\ref{eq:46})) and the solution of the saddle point
equation if $\phi_0(|z|)$.

Thus, it turns out that to bosonize the fermion operators one needs to have an
additional chiral field $X(z)$ as it was discussed e.g. in \cite{AKK,prb}.  In
the following sections~\ref{sec:idls}--~\ref{sec:free-boson} we will remind
this bosonization procedure starting from the quantum mechanics of the
fermions and will show how to blend it with the construction of the
section~\ref{sec:liouville-eq}. In short, the answer will be the following. We
will introduce an additional free scalar field $X(z,\bz)$ in the Liouville
field theory:
\begin{equation}
  \label{eq:48}
  S_\il[X,\phi] =\frac{1}{2 \pi} \int d^2 z \left\{\partial X \bar \partial X
    +  \partial \phi \bar \partial \phi-
    \mu e^{i \phi -| z |^2}
  \right\}
\end{equation}
In terms of the fields $\phi(z,\bz)$ and $X(z,\bz)$ the local fermion operator
in the theory~(\ref{eq:48}) will be defined as
\begin{equation}
  \widetilde\psi^+_\il(\zeta)=\exp\left(\frac {i
      \phi(\zeta,\bar\zeta)+iX(\zeta)-i\bar X(\bar\zeta)}2\right) 
  \label{eq:49}
\end{equation}
where $X(z)$ and $\bar X(\bz)$ are chiral and anti-chiral components of the
field $X(z,\bz)$. 

Expression~(\ref{eq:49}) can be understood as follows.  As we have seen
in~(\ref{eq:46}) $e^{i\phi}$ multiplies the wave function by the factors
$\prod |\zeta-z_i|^2$.  Then, using this fact and eq.~(\ref{eq:47}) (together
with its anti-holomorphic counter-part), we can see that the vertex
operator~(\ref{eq:49}) turns the wave function $\Psi_0(z_1,\dots,z_N)$ into
\begin{eqnarray}
  \widetilde\psi^+_\il(\zeta) \Psi_0(z_1,\dots,z_N)&=&\left(\prod\limits_{i=1}^N
    |z_i-\zeta|^2\,\frac{\zeta - z_i}{\bar
      \zeta - \bz_i}\right)^{\frac12}\Psi_0(z_1,\dots,z_N)\\
  &=& \prod_{i=1}^N (\zeta-z_i) \Psi_0(z_1,\dots,z_N)
  \label{eq:51}
\end{eqnarray}
that is $\widetilde\psi^+$ is indeed local fermion operator!

To summarize, we see that the theory~(\ref{eq:48}) is equivalent to the
quantum mechanics of $N$ non-relativistic fermions in the uniform magnetic
field (including the correct structure of the vertex operators). The relation
between two theories is very similar to the usual bosonization. Indeed, we
know that for relativistic two-dimensional fermions there is an exact
bosonization.  It was shown in~\cite{prb,AKK} that the infinite number of the
non-relativistic fermions in the harmonic (and inverse harmonic) potentials
can be described in terms of free chiral two-dimensional boson.\footnote{This
  bosonization is different from the standard Das-Jevicki
  theory~\cite{das-jev}, because it is not asymptotic in the phase space, but
  is valid \emph{everywhere} and is realized in terms of two dimensional
  \emph{free} boson.} 
Nevertheless, for any large but finite number of particles $N$ such
bosonization is valid in the infra-red regime only. The reason for that is
quite simple. The theory of 2d boson is a relativistic field theory and it has
ultra-violet singularities at small scales.  These singularities are of course
absent in the quantum mechanical system of the finite number of
non-relativistic fermions.  Thus, to have the exact bosonization valid at all
scales, we need more complicated theory which is finite in the ultra-violet.
In case at hand this appears to be the theory~(\ref{eq:48})--(\ref{eq:49}).
Notice, that in spite of the similarities between the actions~(\ref{eq:48})
and that of $c=1$ string theory, they are quite different in nature.  The main
difference is the background term $e^{-\beta |z|^2}$ in the
action~(\ref{eq:48}), which (as discussed in Section~\ref{sec:dens-op-liouv}
after eq.~(\ref{eq:51})) and is responsible for UV regularization of the
theory. As discussed previously, this term cannot be removed by the field
redefinition and thus represents the modification of the theory and not just a
curved background of Liouville theory.

\section{Naive infinite $N$ bosonization}
\label{sec:idls}

In order to derive the action~(\ref{eq:48})--(\ref{eq:49}) we first review
here the derivation of the bosonization procedure for the infinite number of
non-relativistic fermions in the harmonic potential in terms of the free
chiral boson~\cite{c1,prb}.

We begin with the system of $N$ fermions, described by the following ground
state:\footnote{Because all the wave-functions of the LLL are holomorphic
  (apart from the common factor $e^{-\frac12\beta|z|^2}$), we will be omitting
  this factor in the rest of the paper, working in what is known as
  \emph{holomorphic representation}. Thus eq.~(\ref{hall}) is a holomorphic
  representation of the ground state~(\ref{Laughlin}).}
\begin{equation}
  \Psi_0(z_1,\dots, z_N)\equiv\langle z_1,\dots , z_N|
  0\rangle=\frac{1}{\sqrt N!} \prod_{i<j}^{N}(z_j-z_i)^{\beta}
  \label{hall}
\end{equation}
As discussed in Section~\ref{sec:liouville-eq} after eq.~(\ref{coulombH}), the
particle density distribution, described by this function, has the typical
shape presented on the figure~\vref{fig:density}. In this section we are going
to approximate this density by the step-function so that the
state~(\ref{hall}) describes the circular droplet of the radius $\sqrt N$.
The low-lying excitations of the system~(\ref{hall}) are incompressible
deformations (i.e.  those, preserving the total area of the droplet, while
changing its shape). 

To construct a field theoretical description of these degrees of freedom, we
will implement the following strategy~\cite{c1,prb}: first we introduce the
classical sources, describing area-preserving deformations of the
system~(\ref{hall}). Then we will quantize them and construct their Hilbert
space \CH.

Classically, the following set of wave-functions describes any incompressible
deformation of the state~(\ref{hall}):
\begin{equation}
  \langle z_1, \dots, z_N| t_k\rangle=\Psi_0(z_1,\dots,z_N)\,\prod_{i=1}^N
  e^{\beta\omega(z_i)}
  \label{IDLS}
\end{equation}
where $\omega(z)= \sum_{k=1}^{\infty} t_k z^k$ is an entire function. Thus any
state $\ket\tk$ belongs to the LLL.
\begin{figure}[t]
  \begin{center}
    \hbox to \textwidth{\hfil
      \parbox[c]{.3\textwidth}{\includegraphics[scale=.5]{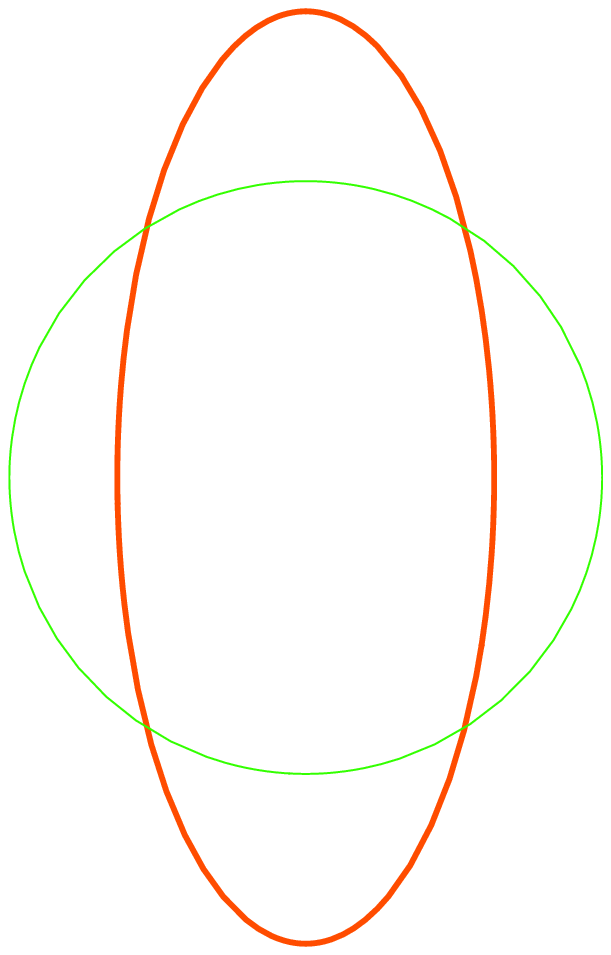}}
      \parbox[c]{.3\textwidth}{\includegraphics[scale=.5]{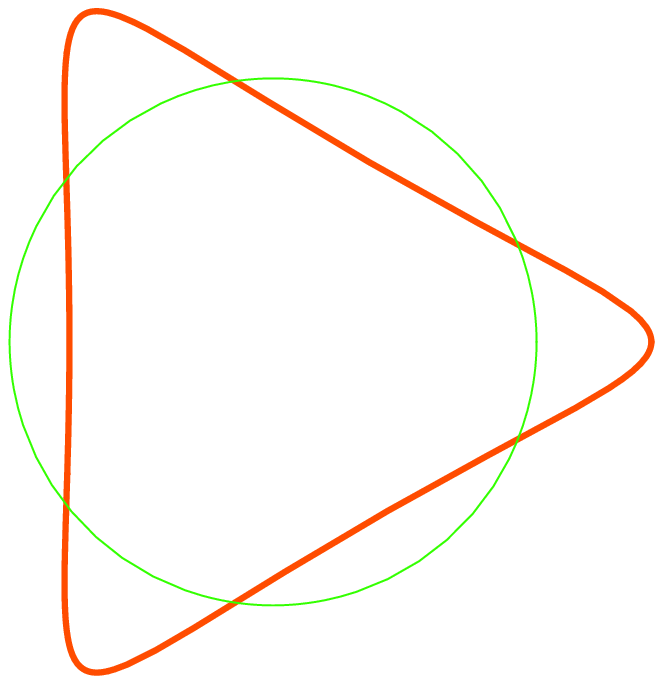}}\hfil}
    \caption{Simplest examples of analytic curves with
      non-trivial \tk's. On the left: ellipse (only $\t2\neq 0$) and on the
      right -- curve with the $\mathbb{Z}_3$ symmetry ($\t3\neq0$). }
    \label{fig:tk}
  \end{center}
\end{figure}
It is easy to understand why these functions describe incompressible
deformations. Density average, computed over the state $\ket\tk$, is constant
inside the contour $\gamma$, that has the same area as the ground
state~(\ref{hall}), and whose shape is parameterized by the set of the so
called \emph{harmonic moments} \tk~(c.f.~\cite{wieg-qhe,prb}):
\begin{equation}
  \label{eq:3}
        \tk \equiv -\frac 1{\pi k}\mathop{\int}\limits_{\mathrm{ext}\;\gamma}
      \frac{d^2z}{z^k};\qquad \tbk \equiv (\tk)^*;\quad k>0
\end{equation}
These moments have simple geometric interpretation.  Circle with the center at
the origin has all harmonic moments equal to zero.  Examples of the simple
shapes with non-zero \tk are given on the figure~\vref{fig:tk}. Under certain
assumptions \tk's serve as coordinates in the space of analytic
curves~\cite{wz,kkmwz}. Thus, at least quasi-classically states~(\ref{IDLS})
clearly describe all possible deformations of the droplet.

On the quantum level, to construct the Hilbert space \CH of incompressible
deformations\footnote{All the deformations we consider in this paper are
  incompressible, therefore we will often call them simply ``deformations of
  the ground state''.}  
of the system~(\ref{hall}) it is convenient to define the so-called
\emph{\trep} (we refer to~\cite{c1,prb} for details) -- isomorphism of Hilbert
space \CH and that of the free chiral boson.

To construct this isomorphism, we consider the matrix element between an
arbitrary state $\ket \Phi\in \CH$ and~$\ket\tk$, which we will call
\emph{\trep of the state $\ket\Phi$}
\begin{equation}
  \Phi(\bar t_k)\equiv\langle t_k| \Phi \rangle =\int d^2 z_1\dots d^2 z_N\,
  \prod_{j=1}^N e^{\beta \bar\omega(\zb j)}\bar\Psi_0(\bz_1,\dots,\bz_N)\Phi(z_1,\dots,z_N)
  \label{trep}
\end{equation}
The scalar product in \trep is defined as
\begin{equation}
  \label{sprod}
  \sprod{\Phi_1}{\Phi_2}_{\bar t} \equiv \bar \Phi_1(\frac1{ k\beta N^k} \frac{\p}{\p\tbk}) \Phi_2(\tbk)\biggr|_{\tbk=0}
\end{equation}
It can be easily shown~\cite{c1,prb} that for any two states
$\ket{\Phi_1},\ket{\Phi_2}\in\CH$ their scalar product in $z$-representation
(i.e.  $\int d^2 z_1\dots d^2 z_N \bar
\Phi_1(\bz_1,\dots,\bz_N)\Phi_2(z_1,\dots,z_N)$) is equal to that defined
via~(\ref{sprod}).  Thus \trep defines an isomorphism of the Hilbert spaces.
The \trep of the ground state~(\ref{hall}) is
\begin{equation}
  \label{eq:5}
  \Psi_0(\tbk) = \int d^2 z_1\dots d^2 z_N\, \prod_{j=1}^N e^{\beta \bar\omega(\zb
    j)}|\Psi_0(z_1,\dots,z_N)|^2 
  =\tau_N(0,\tbk)
\end{equation}
In case $\beta=1$ function $\tau_N(x_k,y_k)$ appearing in~(\ref{eq:5}) is a
tau-function of the 2D Toda lattice hierarchy~\cite{ueno-tak}, while for
$\beta>1$ it lacks interpretation in terms of integrable systems. However, in
the large $N$ limit this function is given by~\cite{kkmwz,wieg-qhe,prb}
\begin{equation}
  \label{eq:6}
  \tau_N(\tk,\tbk) = e^{\beta F(\tk,\tbk)}
\end{equation}
where $F(\tk,\tbk)$ \emph{does not} depend on $\beta$ and is a logarithm of
the tau-function of dispersionless Toda lattice hierarchy
(c.f.~\cite{tak-tak,kkmwz,wz}). Thus, we see that the integrability of the
system is recovered in the large $N$ limit even for $\beta>1$!  An important
property of the function $F(\tk,\tbk)$ is
\begin{equation}
  \label{eq:7}
\left.  \frac{\p^2 F(\tk,\tbk)}{\p t_n \,\p \bar t_m}\right|_{\scriptstyle{\scriptstyle\tk=0} \atop {\scriptstyle\tbk=0}} = n N^n \delta_{n,m}
\end{equation}
with all other first and second derivatives vanishing.

It is easy to define the structure of Fock space in \CH in \trep.  Define
operators
\begin{equation}
  \label{ak}
  a_{-k}=   \sqrt \beta k\:\tbk, \quad
  a_{k} =\frac 1{\sqrt \beta}\frac{\partial}{\partial \bar
    t_k}, \qquad    [ a_{k},\,  a_{-n}]= k \delta_{k,n},\quad k>0
\end{equation}
From the properties~(\ref{eq:7}) of the function $F(\tk,\tbk)$ it follows that
\begin{equation}
  \label{eq:19}
  a_n \ket 0 \Leftrightarrow \frac{\p \tau_N(0,\tbk)}{\p \bar t_n}\Bigr|_{\tk
    = 0} = 0\quad \forall\: n>0 
\end{equation}
i.e. state~(\ref{eq:5}) is a Fock vacuum with respect to the
operators~(\ref{ak}).  States $a_{-k}\ket0$ correspond to the quantized
version of elementary deformations with the harmonic moment \tk (as those on
the figure~\vref{fig:tk}).  In the space \CH there exists a basis of symmetric
polynomials
\begin{equation}
  \ket{ \mathbf n} =\left.
    \frac{1}{\sqrt{\tau_{N}} }\prod_{k=1}^\infty
    \left(\frac{\partial}{\partial t_k}\right)^{\mathbf{n}_k} \ket\tk
  \right|_{t_k=0}.   
  \label{basis}
\end{equation}
For example, state $ \ket{0,\dots ,0,\mathbf{n}_k,0\dots}$ corresponds to
$(\sum\limits_{\scriptscriptstyle j=1}^{\scriptscriptstyle N}
z_j^k)^{\mathbf{n}_k} \Psi_0(z_1,\dots,z_N)$.  Property~(\ref{eq:7}) means
that the states $\ket{\mathbf{n}}$ can be represented as
\begin{equation}
  \label{eq:2}
  \ket{\mathbf{n}} =
  \frac1{\sqrt{\tau_N}}\prod_{k=1}^\infty\frac{a_{-k}^{\mathbf{n}_k}}{(\betas 
    N^k)^{\mathbf{n}_k}} \ket0
\end{equation}
A function $\omega(z)$, entering the definition~(\ref{IDLS}) of an arbitrary
incompressibly deformed state $\ket\tk $, can be expanded in the Taylor series
\begin{equation}
  \label{eq:8}
  \ket\tk = \sum_{\mathbf n} \frac{t_1^{\mathbf{n}_1}\dots
    t_k^{\mathbf{n}_k}\dots}{\mathbf{n}_1!\dots \mathbf{n}_k!\dots} 
 | \mathbf n \rangle
\end{equation}
Combining~(\ref{ak}),~(\ref{eq:19}),~(\ref{eq:2}) and~(\ref{eq:8}) we see that
Hilbert space \CH has the structure of Fock space of the\emph{ free chiral
  boson}:\footnote{Notice, that although for $\beta>1$ ground
  state~(\ref{hall}) corresponds to some interacting system of fermions, the
  theory of its low-lying excitations is still the \emph{free}!}
\begin{equation}
  \label{X}
  X(\zeta)= i\sqrt \beta\sum\limits_{k>0}
  \left[ \frac{a_k}{k\zeta^k} -  \frac{ \zeta^k a_{-k} }{k}      
    \right]- ia_0 \log \zeta  +x_0
\end{equation}
Let us discuss zero modes $a_0$ and $x_0$ of the operator $X(\zeta)$.  We know
that in our theory there should exist a set of vertex operators $V_p(\zeta)$,
which in the first-quantized formalism act as follows:
\begin{equation}
  \label{eq:11}
  V_p(\zeta) \Psi_0(z_1,\dots,z_N) = \prod_{j=1}^N (\zeta-z_j)^p
  \Psi_0(z_1,\dots,z_N),\qquad p=1,\dots,\beta
\end{equation}
In particular we are interested in the operator $V_\beta\equiv \psi^+(\zeta)$
-- \emph{local fermion operator}, that of adding a particle at a point
$\zeta$. It is natural to expect this operator to be realized as
\begin{equation}
  \label{eq:53}
  \psi^+(\zeta) = \no e^{i \sqrt \beta X(\zeta)}\no
\end{equation}
This will indeed be so, if we introduce a zero mode: $a_0=\sqrt\beta N$ and
its conjugated $x_0$: $[a_0,x_0] =1$. Note, that the fact that $a_0$ is
quantized means that the chiral boson is compactified: $X\cong X+ 2\pi R$. Its
radius of compactification can be determined from the quantization condition
for $a_0$ to be $R=\sqrt\beta$ (indeed $X(\zeta e^{2\pi i}) = X(\zeta) + 2\pi
\sqrt\beta N = X(\zeta) + 2\pi R N$).

Let us return to the local fermion operator.  Notice, that we can write
\begin{equation}
  \label{eq:34}
  {}_{N+1}\bra0 \psi^+(\zeta) \ket{0}_N=\zeta^{\beta N}\tau_N\left(0,-\frac\beta{k\zeta^k}\right)
\end{equation}
where ket and bra states in the left hand side should be understood in $N$ and
$N+1$ particle sector correspondingly.  Eq.~(\ref{eq:53}) gives an
\emph{exact} bosonization for infinite number of fermions. By analytic
continuation one can thus show that the $c=1$ MQM also admits \emph{exact}
bosonization in this limit (c.f.~\cite{AKK,c1}).

Finally, from the definition of the scalar product~(\ref{sprod}) it follows
that
\begin{equation}
   (a_k)^+ =N^k\,  a_{-k};\quad a_0^+=a_0,\quad x_0^+ = x_0
  \label{eq:9}
\end{equation}
and thus we can write from the conjugated version of~(\ref{X}):
\begin{equation}
  \label{eq:13}
  \psi(\zeta) = \Bigl[\psi^+(\bar\zeta)\Bigr]^+=\no e^{-i\betas X(\frac
    N\zeta)}\no 
\end{equation}
for the local fermion operator of the annihilation of the fermion at the point
$\zeta$.\footnote{Notice, that strictly speaking the second equality in
  eq.~(\ref{eq:13}) is not true -- it contains some additional $c$-number,
  coming from the correct treatment of the zero modes. In this case there this
  factor is $e^{-\beta N\log N}$.\label{fn:3}} 
Notice, that the expression $\frac N\zeta$ is the point, symmetric to the
point $\bar\zeta$ with respect to the circle with the radius $\sqrt N$
(c.f.~\cite{sft}).

\subsection{Density matrix and UV divergences}
\label{sec:density-matrix}

It is instructive to compute a density matrix to demonstrate the limits of
applicability of the approach of the previous section. The density matrix (or
fermion-fermion correlator) is defined as
\begin{equation}
  \label{eq:14}
  \rho(\bz,w)\equiv\bra 0 \psi^+(\bz) \psi(w)\ket0 = \bra0\no
  e^{i\betas  X(\bar z)}\no\no e^{-i\betas  X(\frac Nw) }\no \ket0
\end{equation}
In computing~(\ref{eq:14}) in \trep, we would have to calculate quite
non-trivial scalar product between the states $\bra\tk\psi(w)\ket0$ and
$\bra\tk\psi(z)\ket0$:
\begin{equation}
  \label{eq:10}
  \bra 0 \psi^+(\bz) \psi(w)\ket0 = (\bz w)^{\beta N}\bar\tau_N(-\frac1{k
    \bz^k},\frac 1{k \beta N^k}\frac{\p}{\p \tbk})\tau_N(-\frac{1}{k
    w^k},\tbk)\Bigr|_{\tbk=0} 
\end{equation}
(factor $(\bz w)^{\beta N}$ comes from the zero mode contribution). Functions
$\tau_N$ are functions of infinitely many variables, which can be defined only
implicitly. Therefore, computation of expression~(\ref{eq:10}) may seem to be
difficult.  However, it is possible to use the following fact.
Recall~\cite{br} that one can interpret the right hand side. of
eq.~(\ref{eq:6}) as the Dirichlet boundary state on the circle of radius
$\sqrt N$:
\begin{equation}
  \label{eq:54}
  \tau_N(\tk,\tbk) = \exp\Bigl(\beta\sum k N^k \tk \tbk + \cdots\Bigr)
  =\exp\Bigl(\sum \frac{a_{k}^+ \bar a_{k}^+}{k  N^k}\Bigr)
\end{equation}
where the last equality is understood in terms of operators
$a_k^+$~(\ref{ak}),~(\ref{eq:9}) and their anti-holomorphic counterparts
(another set of bosonic modes $\bar a_k^+$, realized as in~(\ref{ak}) with
respect to another variable \tk: $\bar a_{-k} = k\tk$, etc.) acting on the
``vacuum'' (constant function).  Then we can rewrite~(\ref{eq:10}) as
\begin{equation}
  \label{eq:18}
  \rho(\bz, w) = \no e^{i\betas (X(\bz) + \bar X(w))}\no \sprod{\tk}{\tk}=\no
  e^{i\betas \left(X(\bz) + \bar X(w)\right)}\no e^{\beta F(\tk,\tbk)}
\end{equation}
From this representation it is trivial to compute that\footnote{In
  eq.~(\ref{eq:17}) we restored factors $e^{-\beta |z|^2}$ and took into
  account additional factor, mentioned in the footnote~\vref{fn:3}.}
\begin{equation}
  \label{eq:17}
  \rho(\bz, w)=e^{-\frac{\beta}2(| w|^2+| z|^2)}
    \frac{(w \bar z)^{\beta N }
    }{N^{\beta N}(1-\frac{N}{ w \bar z})^{\beta}}
\end{equation}
To understand the meaning of this correlator, we can now put $z= \sqrt Nr
e^{i\theta_1}$ and $w=\sqrt N r e^{i\theta_2}$.  Then correlator~(\ref{eq:17})
becomes
\begin{equation}
  \label{eq:4}
  \rho(\theta_1,\theta_2)=\frac{e^{-\beta N(r^2 - \log
      r^2)}}{(1-e^{-i(\theta_1-\theta_2)})^\beta} 
\end{equation}
\begin{figure}[t]
  \centering \includegraphics[scale=.75]{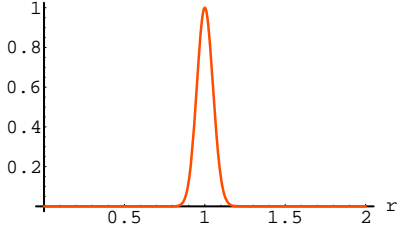}
  \caption{Profile of the fermion correlator in the radial direction. Function
    sharply peaks at $r=1$ and falls off very fast (the case of $\beta N =
    100$ is shown).}
  \label{fig:delta}
\end{figure}
$r$-dependent part of~(\ref{eq:4}) peaks sharply around $r=1$ and goes to zero
both as $r\to0$ and $r\to\infty$ (c.f. Figure~\vref{fig:delta}). On the other
hand, while $|\theta_1 - \theta_2|\ll 2\pi$, one can rewrite the $\theta$-part
as simply $\frac 1{(\theta_1 -\theta_2)^\beta}$.  Thus~(\ref{eq:4}) looks like
a delta-function on the edge of the droplet times the two-point correlator of
field theory of the chiral scalar.

The immediate problem with this correlator is that it is divergent on the
surface of the Fermi-sea (i.e. when $|z|=|w|=\sqrt N$ or $r=1$) as $z\to w$.
We know that this is not the case in the original theory of fermions, as the
answer~(\ref{eq:46}) for $\la\rho(z,\bz)\ra$ and it is finite. The reason for
this bogus divergence has already been discussed throughout the paper: we
tried to bosonize the system of non-relativistic fermions with the
relativistic scalar fields. For this we need strictly infinite number of
fermions. For any arbitrary large but finite number of fermions the procedure,
discussed in the sections~\ref{sec:idls}--\ref{sec:density-matrix} breaks
down, as basis states~(\ref{basis}) become linearly dependent (indeed, for any
finite $N$ there are only finite number of symmetric polynomials of $N$
variables). The other way to see where the approximation breaks down is to see
that the quadratic approximation of tau-function~(\ref{eq:54}) fails for index
$k\sim \sqrt N$ (we refer to Appendix~\ref{sec:expansion}).  Thus, when points
$z$ and $w$ move close enough (distance of the order $1/\sqrt N$) to each
other, the simple theory of the scalar field does not suffice and one needs to
embed it into more general (UV-finite) theory.\footnote{This idea has been
  first emphasized in works~\cite{wadia-w-inf,wadia-time-dep,wadia-beta-func},
  where the UV-finiteness has been achieved essentially by using
  non-commutative field theory.}  As we have already argued in the end of
Section~\ref{sec:liouville} correct short-distance limit for the density
correlators can be achieved on the field-theoretical side by using (modified)
Liouville theory.  Below we are going to combine two approaches (that of the
Section~\ref{sec:liouville} with that of the current section) to construct the
correct fermion operators and derive the theory~(\ref{eq:48})--(\ref{eq:49}).

\section{Imaginary Liouville theory plus free boson}
\label{sec:tk}

Our strategy during the next steps of the derivation of the
eqs.~(\ref{eq:48})--(\ref{eq:49}) will be to apply the construction of
Section~\ref{sec:idls}, using for the densities and the normalization $\tau_N$
the field theoretical representation in terms of Liouville field found in
Section~\ref{sec:liouville}.  To do so we need to generalize this
representation for the $\tau_N(\tk,\tbk)$, i.e. for the case of an arbitrary
incompressibly deformed background $\ket\tk$:
\begin{equation}
  \langle z_1, \dots , z_N\ket\tk=\frac1{\sqrt{N!}}\prod_{i<j}^N
  (z_i-z_j)^{\beta} 
  e^{\left[\beta\sum\limits_{j=1}^N 
    \left(\omega(z_j)-\frac{| z_j|^2}{2}\right)\right]}
\label{Laughlin2}
\end{equation}
with $\omega(z)$ as in~(\ref{IDLS}).  The norm of this state is written as
\begin{equation}
  \label{eq:30}
  \tau_N(\tk,\tbk)\equiv \sprod{\tk}{\tk} = \int d^2 z_1 \dots d^2 z_N
  e^{-\beta \CE_\omega}
\end{equation}
where
\begin{equation}
  \label{eq:41}
  \CE_\omega =-2\sum_{i<j}  \log| z_i-z_j| -\sum_{j=1}^{N}
\Bigl[\omega(z_j)+\bar \omega(\bar z_j)-| z_j|^2 \Bigr]
\end{equation}
Going to the grand canonical ensemble, we arrive to the following modification
of the theory
\begin{equation}
  \CZ(t_k, \bar t_k, \mu)=\sum_{N=0}^{\infty}
  \frac{\xi^N}{N!}\langle t_k| t_k \rangle_N =
  \frac{1}{Z_{\phi}}
  \int \CD \phi\, e^{-S_\il[\phi, \tk, \tbk]}
  \label{Zoft}
\end{equation}
where
\begin{equation}
  \label{eq:27}
  S_\il[\phi, t_k, \bar t_k]=\frac{1}{2 \pi\beta} \int d^2 z
  \left\{\partial \phi \bar \partial \phi-
     \mu\, e^{i \phi + \beta[\omega(z)+\bar \omega(\bar z)]
      -\beta | z |^2}
  \right\}
\end{equation}
with $\mu$ is given by the same expression~(\ref{eq:26}).  It is convenient to
use a slightly different representation of the same action.  Consider the
following shift
\begin{equation}
  \label{eq:55}
  \phi(z, \bar z) \to \phi(z, \bar z) +i\beta [\omega(z)+\bar \omega(\bar z) ]
\end{equation}
Then the action becomes
\begin{equation}
  S_\il[\phi, t_k, \bar t_k]=\frac{1}{2 \pi\beta} \int d^2 z
  \left\{\partial \phi \bar \partial \phi-
    \mu\, e^{i \phi  -\beta| z |^2}
  \right\}+ S_{b}
  \label{Soft}
\end{equation}
However, the shift~(\ref{eq:55}) changes the boundary conditions for the field
$\phi$. New field does not vanish at $z=\infty$, but rather has there a
singularity (that of the entire functions $\omega(z)+\bar\omega(\bz)$).  These
boundary conditions can be incorporated in a form of the boundary term
\begin{equation}
  S_{b}=\frac{1}{4\pi}\left[\oint d z\, \phi\, \partial \omega  -
    \oint d\bar z\, \phi \,\bar \partial \bar \omega\right]  -
  \frac{\beta}{8\pi i } \left[\oint dz\, \bar \omega \partial \omega - \oint
    d\bz\, \omega \bar\p\bar \omega\right]
  \label{BT}
\end{equation}
where the integrals are taken around $z=\infty$.

\subsection{Coupling with the free boson}
\label{sec:free-boson}

As a starting point, notice that the boundary terms in~(\ref{BT}) look like
classical sources (given by $\p \omega(z)$ and $\bar \p \bar\omega(\bz)$)
coupled to the field $\phi(z,\bz)$. In the logic of the
sections~\ref{sec:idls}--\ref{sec:density-matrix} we may try to ``quantize''
these sources, by promoting them to the creation operators of some field.
Essentially, we have already done that in eq.~(\ref{ak}), but we have two sets
of variables (\tk and \tbk) in eqs.~(\ref{eq:27})--(\ref{BT}), which means
that the whole scalar field $X(z, \bz)$ should appear. 

To begin with, we compute average of the vertex operator of the Liouville
field:
\begin{equation}
  \label{eq:12}
  \la e^{i\alpha\phi(\zeta,\bar \zeta)} \ra_{t_k} = \frac 1{Z_\phi}\int \CD\phi\,
  e^{-S_\il[\phi,\tk,\tbk]} e^{i\alpha\phi(\zeta,\bar \zeta)} =
  \CZ\left(\tk-\frac 
  \alpha{k\zeta^k},\tbk -\frac \alpha{k \bar\zeta^k},\mu\right)
\end{equation}
where $\alpha$ is an arbitrary number and $\CZ$ is defined by~(\ref{Zoft}).
Careful treatment of the grand canonical ensemble (GCE) in our case shows that
all the correlators, computed in the GCE and in the micro-canonical ensemble
with the $N=\la N\ra_{GCE}$ coincide (as $\bar N\to \infty$) (for details see
Appendix~\ref{sec:careful} and~\cite{liouv-big}). Thus, we can simply
substitute the right hand side of eq.~(\ref{eq:12}) with $\tau_N(\tk-\frac
\alpha{k\zeta^k},\tbk -\frac \alpha{k \bar\zeta^k})$, understanding $N$ as a
function of chemical potential $\mu$ in a sense just
described. 
Namely, if we introduce two sets of Heisenberg operators (as in
Section~\ref{sec:density-matrix}):\footnote{Throughout this section we have
  chosen to ignore zero mode contributions. This is done to keep the
  presentation as simple as possible and avoid proliferation of many
  non-essential factors. It should be straightforward to restore all the zero
  modes contributions.}
\begin{eqnarray}
  \label{osc}
  &&a_{-k}=   \sqrt \beta k\:\tbk, \quad
  a_{k} =\frac 1{\sqrt \beta}\frac{\partial}{\partial \bar
    t_k}, \qquad    [ a_{k},\,  a_{-n}]= k \delta_{k,n},\quad k>0\\
   && \bar a_{-k}=   \sqrt \beta k\:\tk, \quad
  \bar a_{k} =\frac 1{\sqrt \beta}\frac{\partial}{\partial \tk}, \qquad
  [\bar a_{k},\,  \bar a_{-n}]= k \delta_{k,n},\quad [a_k, \bar a_n]
  =0\nonumber 
\end{eqnarray}
define the corresponding scalar fields (compare with eq.~(\ref{X})):
 \begin{gather}
   X(\zeta)= i\sqrt \beta\sum\limits_{k>0} \left[ \frac{a_k}{k\zeta^k} -
     \frac{ \zeta^k a_{-k} }{k}
   \right]+\text{zero modes}\\
   \bar X(\bar\zeta)= i\sqrt \beta\sum\limits_{k>0} \left[ \frac{\bar
       a_k}{k\bar \zeta^k} - \frac{\bar \zeta^k\bar a_{-k} }{k}
   \right]+\text{zero modes}\notag
\end{gather}
and denote by $\ket\vac $ the Fock vacuum of~(\ref{osc}): $ a_k \ket\vac=0$,
$\bar a_k \ket\vac=0$ for $k>0$, then one naturally defines a \emph{t-\trep}
(as a straightforward generalization of construction of
Section~\ref{sec:idls}). Now we may interpret $\tau_N(\tk-\frac
\alpha{k\zeta^k},\tbk -\frac \alpha{k \bar\zeta^k})$ as a state $\ket{\tau_N}$
in the Hilbert space $\CH\otimes \bar\CH$ in $t$-\trep:
\begin{equation}
  \label{eq:32}
  \sprod{\tk,\tbk}{\tau_N} =\tau_N(\tk-\frac \alpha{k\zeta^k},\tbk -\frac \alpha{k
  \bar\zeta^k}) = 
\bra{\tk,\tbk}\tau_N\left(a^+_k-\frac \alpha{k\zeta^k},\bar
  a^+_k-\frac \alpha{k \bar\zeta^k}\right)\ket\vac
\end{equation}
Define the the vertex operator
\begin{equation}
  \label{eq:31}
 \widetilde \psi^+_X(\zeta,\bar\zeta)=\no  e^{\frac i2\betas \left(X(\zeta)
     -\bar X(\bar \zeta)\right)}\no
\end{equation}
In $t$-\trep it acts on the function~(\ref{eq:32}) by shift in both \tk and
\tbk:
\begin{equation}
  \label{eq:16}
  \bra\vac\widetilde\psi^+_X(\zeta,\bar\zeta)\ket{\tau_N}=\tau_N\left(\frac
    \beta{2k \bar\zeta^k}-\frac 
  \alpha{k\bar\zeta^k},-\frac \beta{2k \zeta^k} 
  -\frac \alpha{k \zeta^k}\right)
\end{equation}
If one chooses $\alpha=\frac\beta 2$, then eq.~(\ref{eq:16}) is equivalent to
\begin{eqnarray}
  \label{eq:35}
  \bra\vac\no e^{\frac i2\betas \left(X(\zeta) -\bar X(\bar
      \zeta)\right)}\no\ket{\tau_N}&=&
  \tau_N\left(0,-\frac\beta{k\zeta^k}\right)={}_{N+1}\bra 0
  \psi^+(\zeta)\ket{0}_N\\ 
  &=&
\int d^2 z_1\dots d^2 z_N \bar \Psi_0(\zb 1,\dots,\zb N)\psi^+(\zeta)
\Psi_0(z_1,\dots,z_N)  \nonumber
\end{eqnarray}
i.e. eq.~(\ref{eq:35}) indeed describes the average of the local fermion
operator.

Combining the results of~(\ref{eq:12}),~(\ref{eq:32}),~(\ref{eq:31})
and~(\ref{eq:35}), we arrive to the conclusion that the vertex operator
\begin{equation}
  \label{eq:20}
  \boxed{%
    \vphantom{\biggl|}\widetilde\psi^+(\zeta, \bar \zeta)= e^{\frac{i}{2}\betas[ X(\zeta)- \bar
    X(\bar \zeta)]+  \frac i2\phi(\zeta, \bar \zeta) } 
}%
\end{equation}
is the local fermion operator in imaginary Liouville theory, coupled to the
free scalar:
\begin{equation}
  \label{eq:40}
  \boxed{%
    \vphantom{\Biggl|} S_\ilx[X, \phi]=\frac{1}{2 \pi} \int d^2 z
    \left\{\partial X \bar \partial X 
    + \frac1\beta \partial \phi \bar \partial \phi-
    \frac{\mu}\beta e^{i \phi -\beta| z |^2}
  \right\}+S_{b}
}%
\end{equation}
where the only interaction between the fields $\phi(z,\bz)$ and $X(z,\bz)$ is
through the boundary terms, as in~(\ref{BT}):
\begin{equation}
  S_{b}=\frac{i\betas}{4\pi}\left[\oint d z\, \phi\, \partial X - \oint d\bar
    z\, \phi \,\bar \partial  X\right] + \frac{\beta^2}{8\pi i }\left[ \oint
    dz\, X \partial X-\oint d\bz \, X\bar\p X\right]
    \label{eq:1}
\end{equation}

Several comments are in order here. First, we could have written the same
expression entirely in the path integral formalism, where it would look like
\begin{equation}
  \label{eq:29}
  \la\dots \widetilde\psi^+(\zeta,\bar\zeta)\ra =\int \CD X e^{-S_0[X]}\dots
  e^{\frac{i}{2}\betas[ X(\zeta)- \bar X(\bar \zeta)]}\int \CD \phi\,
  e^{-S_\il[\phi,\tk,\tbk]} e^{\frac i2 \phi(\zeta,\bar \zeta)}
\end{equation}
(where $S_0[X]$ is an action of free massless scalar field $X$ and \tk, \tbk
are the positive Fourier modes of the field $X$).  In the functional integral
over $\phi$ positive frequency modes of the field $X$ enter only via the
boundary terms~(\ref{BT}). One can perform the functional integration over
$\phi$ to obtain a tau-function $\tau_N(\tk,\tbk)$.  The remaining integral
over $X$ with such insertion is equivalent to the operator
computations~(\ref{eq:16})--(\ref{eq:35}).

One can understand the meaning of this insertion as follows. In operator
formalism tau-function $\tau_N$ is an element of the Fock space of the field
$X$, created by $\tau_N(a_k^+,\bar a_k^+)\ket\vac$. If one uses the naive
$N\to\infty$ limit and keeps only quadratic in $\tk$ part of $\tau_N$, one can
regard state $\ket{\tau_N}$~(\ref{eq:32}) as a Dirichlet boundary state on the
circle with the radius $\sqrt N$~\cite{br}:
\begin{equation}
  \label{eq:38}
    \tau_N(\tk,\tbk) = \exp\Bigl(\beta\sum k N^k \tk \tbk + \cdots\Bigr)
  =\exp\Bigl(\sum \frac{a_{k}^+ \bar a_{k}^+}{k  N^k}\Bigr)\ket\vac
\end{equation}
(compare with Section~\ref{sec:density-matrix}). Then eq.~(\ref{eq:35})
becomes
\begin{equation}
  \label{eq:39}
  \bra\vac\dots e^{\frac{i}{2}\betas[ X(\zeta)-
  \bar X(\bar \zeta)]}\bket{\mathrm{D}} = \bra0\dots e^{i\betas
  X(\zeta)}\ket0
\end{equation}
In this approximation we recover bosonization of Section~\ref{sec:idls} and
the theory~(\ref{eq:40}) reduces to that of the free chiral scalar in the
boundary state formalism.

\section{Comparison with the $c=1$ string theory}
\label{sec:2dstr}

As it was discussed in the introduction, although the theory we derived
resembles very much the usual $c=1$ string theory, it has very different
properties and interpretation. First of all, instead of Liouville field it
contains that of with imaginary charge. One could argue that if we repeat this
derivation for the inverse harmonic potential, we could get usual Liouville
field.  To do this we need to continue analytically from $\phi$ to $i \phi$.
Naively if one does such a continuation, the kinetic term in the action
changes its sign and the action becomes unbounded from below. However, as it
was argued in~\cite{c1}, to relate harmonic and inverse harmonic potential
cases, one should make also analytic continuation in $z$ plane to get real
variables $z_+$ and $z_-$ instead of complex (conjugated) variables $z$ and
$\bar z$.  Then the action becomes Lorentzian and the analytic continuation in
$\phi$ does change its properties. Another problem is related to the behavior
of the potential. If we consider its dependence on the constant $\phi$, in the
complex plane there is a line where the potential is
unbounded~\cite{analyt-cont}.  This may suggest that two theories are not
equivalent and the derivation for the inverse harmonic potential should be
done independently.  Generalization of our approach to this case should be
possible and we plan to do it in the future.

Another important feature of our theory is that it is gives, by construction,
finite correlators for the vertex operators, corresponding to the fermions in
the quantum mechanical picture. This property, natural for the effective
theory of non-relativistic fermions, looks strange from field-theoretical
point of view.  It cannot be present in the CFT. In fact, our theory is not
conformal because of the presence of an unusual background term. Modified
Liouville field in our theory is in some sense auxiliary. If one integrates it
out, one gets an effective action for the field $X$. This action is
complicated because it should give UV finite correlators of the vertex
operators. If one uses the simplest approximation for the integral over the
Liouville field (as we discussed at the end of Section~\ref{sec:free-boson}),
one gets just a sharp wall -- Dirichlet boundary state making $X$ chiral.
However, this approximation is non-homogeneous in $z$-plane, as it was
discussed above (Section~\ref{sec:liouville}).  Indeed, near the boundary of
the droplet (Fermi sea) the step function, which is fermion density,
corresponding to the naive large $N$ approximation, is always far from the
real density profile (in a sense that for any $N$ there is $z$ such that the
difference between the exact $\rho_N$ and the step function is of order
unity).  If one uses more careful approximation for the Liouville field, the
effective action of the $X$ becomes more complicated after we integrate out
modified Liouville field.

Thus what we have obtained a UV-finite theory, describing bosonization of free
fermions. Because of its apparent similarity with the two-dimensional string
theory, it is natural to ask: ``is this some sort of string theory?''. The
answer is the following: if there is a string theory description of our
theory~(\ref{eq:40}), it is a string theory in its target space formulation, a
string field theory. Then it is interesting to understand what should be the
world-sheet description of such a theory, if it exists.\footnote{When this
  paper was finished, the work~\cite{izh-mcg} appeared, which suggested a
  string theory, which can serve as a possible candidate for the world-sheet
  description of our theory. A remarkable feature of the theory~\cite{izh-mcg}
  is that its world-sheet action is quite similar to that of~(\ref{eq:40}).
  The relation between the two theories is currently under investigation.} 
It would be also interesting to see how such a description will look like for
the inverse harmonic potential. As we discussed above, we expect the
derivation to be very similar and the answer to differ from the case at hand
by real instead of imaginary Liouville field and Lorentzian signature.  Other
features of our theory should remain unchanged. It should be possible to
compare the action which appears after integrating out the Liouville field
with the Das-Jevicki action.

\subsection*{Acknowledgments}
\label{sec:ackn}

We acknowledge useful communications with A.~Abanov, S.~Alexandrov,
V.~Kazakov, I.~Kostov, E.~Martinec, N.~Nekrasov, J.~McGreevy, H.~Verlinde,
S.~Wadia, P.~Wiegmann, A.~Zabrodin.

\appendix

\section{Careful treatment of the ultra-violet and infra-red regulators}
\label{sec:careful}

In this section we carefully treat both IR and UV regularizations of the
theory~(\ref{S0}).  For the grand canonical ensemble of the Coulomb
plasma~(\ref{eq:22})
\begin{equation}
  \label{eq:23}
  \CZ(\xi)  \equiv \sum_{N=1}^\infty \frac{\xi^N}{N!}\tau_N=\sum_{N=1}^\infty
  \frac{\xi^N}{N!} \int d^2 z_1 \dots d^2 z_N |\Psi_0(z_1,\dots,z_N)|^2
\end{equation}
we rewrite the Hamiltonian~(\ref{coulombH}) in terms of a correlator in a free
boson field theory:
\begin{equation}
  e^{-\beta\CE_{c,reg}(z_1, \dots, z_N)} =\left \langle \prod_{j=1}^N e^{i \phi(z_j,
      \bar z_j)-\beta | z_j|^2} \right \rangle_{S_m}
  \label{HregFT}
\end{equation}
where now the average is taken with respect to the Euclidean Gaussian action
of a free \emph{massive} boson (thus introducing infra-red regulator into the
theory):
\begin{equation}
  S_m[\phi]=
  \frac{1}{2 \pi \beta} \int d^2 z
  \left[\partial \phi \bar \partial \phi+m^2 \phi^2 \right]
  \label{S0m}
\end{equation}
with a short-distance cutoff $\Lambda$ (see Eq.~(\ref{bosecorf}) below)
implied. 
Formula (\ref{HregFT}) is exact and follows from the the Gaussian
integration formula
\begin{equation}
\left\langle\prod_{j=1}^N
e^{i \phi(z_j, \bar z_j)} \right\rangle_{S_m}=
\exp\left[-\frac{1}{2}\sum_{i,j=1}^{N} \langle \phi(z_i, \bar z_i)
\phi(z_j, \bar z_j) \rangle_{S_m} \right]
\label{Gaussian}
\end{equation}
where the correlation function of the field $\phi$ is
\begin{equation}
\langle \phi(z, \bar z) \phi(0) \rangle_{S_m}=\left\{
\begin{array}{rl}
\displaystyle
2\beta K_0(m | z |), & \qquad\Lambda
| z| >1 \vspace{5 pt} \\
\displaystyle 2\beta \ln\left(\frac{\Lambda}{m}\right).
& \qquad\Lambda | z| <1
\end{array} \right.
\label{bosecorf}
\end{equation}
Here $K_0(x)$ is the MacDonald function with asymptotics given
in Eq.~(\ref{MC}).
\begin{equation}
K_0(x)=\left\{
\begin{array}{rl}
-\ln x -\gamma+\ln 2  + O(x), & \qquad x\to 0, \vspace{5pt}\\
\sqrt \frac{\pi}{2 x} e^{-x}[1+o(1/x)], & \qquad x \to \infty
\end{array}
\right.
\label{MC}
\end{equation}
For the regularized version of the partition function $Z_N(\beta)$,
eq.~(\ref{PF}), the representation (\ref{HregFT}) gives
\begin{equation}
Z_{N,reg}(\beta)=\left(\frac{\Lambda}{ m}\right)^{\beta N} \int d^2
z_1\dots d^2 z_N \left \langle \prod_{j=1}^N e^{i \phi(z_j, \bar
z_j)-{\beta} | z_j|^2} \right \rangle_{S_m}
\label{ZNFTreg}
\end{equation}
From the asymptotic~(\ref{MC}) one finds that
\begin{equation}
  \label{eq:36}
  \tau_N = \lim_{m\to 0}\left(\frac{m^2 e^{2\gamma}}{4}\right)^{-\frac\beta2
    N(N-1)} Z_{N,reg}(\beta)
\end{equation}
where $\tau_N$ is given by~(\ref{ZN}). A reason for having the infinitesimal
mass $m$ in the action~(\ref{S0m}) is clear from eq.~(\ref{Gaussian}). For
exactly zero mass the integration over the zero mode of field $\phi$ would
make the right hand side exactly zero.

Substituting the representation~(\ref{ZNFTreg}) in~(\ref{eq:23}) we find
\begin{equation}
\CZ(\xi)=\frac{1}{Z_{\phi}} \sum_{N=0}^{\infty} \frac{\xi^N}{N!} \left(\frac{\Lambda}{m} \right)^{\beta N} \int \mathcal D\,
\phi e^{-S_m} \left[\int d^2 z e^{i \phi
-{\beta} | z|^2 }\right]^N. \label{Xisum}
\end{equation}
The sum over $N$ is easily performed yielding
\begin{equation}
\CZ(\mu) = \frac{1}{Z_\phi} \int \mathcal D \phi e^{-S_{\il,reg}[\phi]},
\label{Xiint}
\end{equation}
where
\begin{equation}
S_{\il,reg}[\phi]= \frac{1}{2 \pi \beta} \int d^2 z \left[\partial \phi \bar
\partial \phi+m^2 \phi^2 -\mu e^{i \phi- \beta |
z|^2 }\right] \label{SLio}
\end{equation}
Equations (\ref{Xiint}) and (\ref{SLio}) define the path integral
representation for $\CZ$. The coupling constant
\begin{equation}
\mu = \left(\frac{\Lambda}{m} \right)^{\beta}
 2\pi \beta \xi
 \label{xidef}
\end{equation}
has the dimension of inverse length squared. It is a running
coupling constant satisfying the RG equation
\begin{equation}
\frac{d \log\mu}{d\log \Lambda}= \beta
\end{equation}

We are interested in the $m\to 0$ limit of the grand partition
function~(\ref{eq:23}). In this limit grand partition function~(\ref{eq:23})
possesses a remarkable property: grand canonical ensemble gives an
\emph{exact} approximation for the averages calculated in the microcanonical
ensemble.

Indeed, assuming $\bar N\gg 1$ (where $\bar N$ is a fixed average number of
particles) one can perform the sum in eq.~(\ref{Xisum}) in the saddle point
approximation. The saddle point equation defines the chemical potential
$\chi=\log\xi$ in terms of the average particle density $\bar N$
\begin{equation}
\chi=\frac{d}{d\bar N} \Omega_N - \bar N
\ln \frac{m^2 e^{2\gamma}}{4}+ \frac{1}{\beta} \log\bar N
\label{mua}
\end{equation}
where $\Omega_N$ is a thermodynamic potential of the Coulomb
plasma~(\ref{PF}). For $\bar N\gg 1$ it has the following asymptotic (see
e.g.~\cite{kkmwz}):
\begin{equation}
  \label{eq:33}
  \Omega_N = - \frac{1}{2} N^2 \ln N +\frac{3}{4} N^2 +W(N)
\end{equation}
This can be understood by computing the free energy of the uniformly charged
disk of radius $\sqrt N$ in the uniform background of the unit charge (the subleading
term $W(N)=\CO(N\ln N)$ is the correlation energy).  The fluctuation of the
number of particles $\delta N$ around the saddle point is given by the
standard thermodynamics formula
\begin{equation}
  (\delta N)^2 = \frac{1}{\beta}  \frac{d \bar N}{d \chi}
\end{equation}
Substituting in this equation the result eq.~(\ref{mua}) for the chemical
potential $\chi$ and using formula (\ref{eq:33}) we find
\begin{equation}
  (\delta N)^2=\left(\log \frac{\mathrm{const}}{\bar N m^2}\right)^{-1}.
  \label{deltaN}
\end{equation}
This is a remarkable formula. It suggests that in the limit $m\to 0$ the
number of particles in the plasma does not fluctuate and the grand canonical
ensemble gives an exact approximation for the averages calculated in the
microcanonical ensemble.

Last in this section we compare the average calculated in
canonical and the grand canonical ensemble. Consider an
observable $\mathcal O(z_1, \dots, z_N).$ Denote its average in the
microcanonical ensemble
\begin{equation}
\mathcal O_N= \frac{1}{Z_{N,reg}(\beta)} \int d^2 z_1 \dots d^2 z_N\, e^{-\beta
\mathcal \CE_{c,reg}} \mathcal O(z_1, \dots, z_N)
\end{equation}
The average in the Grand canonical ensemble will then
be given by
\begin{equation}
  \langle \mathcal O \rangle =\frac{1}{\CZ} \sum_{N=0}^\infty \frac{\xi^N}{N!}
  Z_{N,reg}(\beta) \mathcal O_N 
  \label{grandav} 
\end{equation}
Performing the sum in the saddle point approximation we find
\begin{equation}
\langle \mathcal O \rangle  = \mathcal O_N+
(\delta N)^2 \frac{1}{2} \frac{d^2}{d N^2} \mathcal O_N.
\end{equation}
By virtue of Eq.~(\ref{deltaN}) the correction vanishes
in the limit $m\to 0:$
\begin{equation}
\langle \mathcal O \rangle \rightarrow \mathcal O_N , \qquad m\to0
\label{exactness}
\end{equation}

\section{Expansion of the density}
\label{sec:expansion}

In this section it will introduce an additional parameter $h=1/N$ to keep the
area of the droplet of the ground state finite and keep $\beta=1$ for
simplicity.  The asymptotic expression for the one-particle density matrix is
obtained from the exact formula~(\ref{eq:10})
\begin{equation}
\rho(\zeta, \bar \zeta)=
e^{-\frac{1}{h}\vert \zeta \vert^2}
\frac{1}{\tau_{N+1}(0)}\zeta^{N+1} \bar \zeta^{N+1}\tau_N \left(
\textstyle
-\frac{h}{k \zeta^k}, -\frac{h}{k \bar \zeta^k}
\right)
\label{IGcorf}
\end{equation}
Using the small $t_k$ expansion for the logarithm of the $\tau$-function
\begin{equation}
\log \frac{1}{\tau_{N+1}(0)}\tau_N(t_k, \bar t_k)= \log C_N+ \frac{1}{h^2}
\sum_{k\ge 1} k t_k \bar t_k+ o(\vert t_k \vert^3).
\label{IGlogtau}
\end{equation}
(where $C_N=\frac{1}{N!h^{N+1}}$), for the $\tau$-function in (\ref{IGcorf})
we find
\begin{equation}
\frac{1}{\tau_{N+1}(0)}\tau_N\left(
\textstyle
-\frac{h}{k \zeta^k}, -\frac{h}{k \bar \zeta^k}
\right) \simeq\frac{C_N}{1-(\zeta \bar \zeta)^{-1}} = C_N \sum_{k\ge 1}\frac{1}{\zeta^k\bar \zeta^k}
\label{IGasy}
\end{equation}
It is instructive to compare the asymptotic formula (\ref{IGasy})
with the exact result in the $\beta=1$ case, which can be found from
comparing the expression (\ref{IGcorf}) with the exact density matrix
of free fermions
\begin{equation}
\rho(\zeta, \bar \zeta)=e^{-\frac{1}{h}\vert \zeta \vert^2} \sum_{k=0}^{N}
\frac{\zeta^k \bar \zeta^k}{h^{k+1}k!}.
\label{IGexact}
\end{equation}
The resulting expression for the $\tau$-function reads
\begin{equation}
\frac{1}{\tau_{N+1}(0)}\tau_N\left(
\textstyle
-\frac{h}{k \zeta^k}, -\frac{h}{k \bar \zeta^k}
\right) = C_N \sum_{k=0}^{N} \frac{N!}{(N-k)!}\frac{h^k}{\zeta^k \bar\zeta^k}
\label{IGtauexact}
\end{equation}
One can see that for $h=1/N$ the coefficients of the sum (\ref{IGtauexact}) are
rapidly decaying with $k.$ Simple asymptotic analysis shows that for small
$k$ the Stirling formula can be used for all factorials giving after some
algebra
\begin{equation}
 \frac{N!h^{k}}{(N-k)!} \sim \exp\left[-N \left(\frac{k^2}{1\cdot 2 N^2}+ \frac{k^3}{2\cdot 3 N^3}+
 \dots \right) \right].
\end{equation}
One can see that this expression is exponentially small for $p>\sqrt N$ and therefore
only the leading term in the exponential can be kept. This gives
\begin{equation}
\frac{1}{\tau_{N+1}(0)}\tau_N\left(
\textstyle
-\frac{h}{k \zeta^k}, -\frac{h}{k \bar \zeta^k}
\right) \simeq C_N \sum_{k=0}^{N} e^{-\frac{1}{2} \frac{k^2}{N}} {\frac{1}{\zeta^k \bar\zeta^k}}
\label{IGexas}
\end{equation}
Comparing equation (\ref{IGexas}) with (\ref{IGasy}) one can see that the
asymptotic expansion (\ref{IGasy}) breaks down for $p\sim \sqrt N.$ The same
is true for the small $t_k$ expansion (\ref{IGlogtau}).

\end{document}